# The Composite Particles Model (CPM), Vacuum Structure and ~ 125 GeV Higgs Mass

Marko B. Popovic*

May 15, 2012

**Abstract**
The Composite Particles Model (CPM) is characterized by composite Higgs, composite top quark, cancelation of the scalar leading quadratic divergences, and a particular ground state such that top anti-top channel is neither attractive or repulsive at tree level at the Z pole mass. The radiatively generated scalar mass in 2D is $m_H = \sqrt{\frac{6m_t^2 - M_Z^2 - 2M_W^2}{3\left(1+\frac{\pi}{k}\right)}} = 113 \text{ GeV}/c^2, 143 \text{ GeV}/c^2, \ldots, 230 \text{ GeV}/c^2$ for k = 1, 2, ... ∞ .
As first proposed by Nambu, in the simplest models with dynamical mass generation and fermion condensate in 4D, one expects the Higgs mass on the order of twice the heaviest fermion mass. Hence, if this is applied to the CPM one could expect scalar mass dynamically generated by top constituent quarks and composite top quarks to be equal to $2\frac{m_t}{3}$ and $2m_t$ respectively. When Bose-Einstein statistics for $kT \cong M_W c^2$ is applied to the two lowest energy states in 2D (113 GeV and 143 GeV) and 4D (115 GeV and 346 GeV), the CPM suggests physical Higgs mass equal to $m_H \cong 125 \text{ GeV}/c^2$ in both 2D and 4D.

*Please Address All Correspondence To:
Marko B Popovic, Physics Department, Worcester Polytechnic Institute, 100 Institute Road, Worcester, MA 01609, USA Cell: 617 470 8198, Email: mpopovic@wpi.edu

## 1. Introduction

The molecules, atoms, nuclei, baryons, mesons, etc. are all composite particles. While these particles may appear elementary in the infrared their composite nature is readily revealed in the ultraviolet. The compositeness can be sometimes even inferred from the physics in the infrared. The CPM [1-3], analyzed here, extends this "compositeness principle" by encompassing dynamical electroweak symmetry breaking (EWSB) with composite Higgs and composite top quark. While numerous recent studies [4] address composite Higgs none has addressed possibility that top quark may be also a composite particle.

The Higgs mechanism with a weak doublet scalar field having a non-zero vacuum expectation value (VEV) provides the simplest potential realization of the spontaneous EWSB. However, if Higgs is an elementary particle, a natural explanation of potentially large hierarchy might be difficult. The renormalization group running of the SM scalar effective mass squared $m^2_{H\,eff}(\Lambda^2)$ is quadratic function of energy scale, $\Lambda$., hence, a small change at one end of the energy spectrum is traditionally assumed to lead to tremendous changes at the opposite end of the energy spectrum. This quantity can be made dimensionless and represented as $\mu = \frac{m^2_{H\,eff}(\Lambda^2)}{\Lambda^2}$. If EWSB is dominated by the scalar sector the EWSB scale, $\Lambda_{EWSB}$, is expected to correspond to an energy scale at which the dimensionless effective Higgs mass squared changes its sign [5,2], i.e. $\mu(\Lambda^2_{EWSB}) = \frac{m^2_{H\,eff}(\Lambda^2)}{\Lambda^2}|_{\Lambda=\Lambda_{EWSB}} = 0$ or alternatively $m^2_{H\,eff}(\Lambda^2_{EWSB}) = 0$ The detailed mapping between possible physical Higgs masses, $m_{H\,phys}$, and $\Lambda_{EWSB}$, has been obtained independently by two standard regularization schemes: hard-cutoff and $\overline{MS}$, [5,2]. The $\Lambda_{EWSB}$ is obtained to be in the range of 800-1050 GeV for physical Higgs masses in the range of 115-145 GeV. If the SM is good approximation of theory in the infrared, then this mapping is expected to persist even if the Higgs is not an elementary particle [3].

For experimentally preferred physical Higgs masses, the dimensionless effective mass squared in ultraviolet, i.e. $\mu(\Lambda^2)$ for $\Lambda > \Lambda_{EWSB}$, is a smooth, slowly varying function all the way to the Planck scale; however, the vacuum stability constrains the maximally allowed scale to be less than the Planck scale [5,2]. This function resembles a suppressed $\frac{d\mu}{dln(\Lambda)} \propto -\mu \ll 1$ mode. In the infrared, $\Lambda < \Lambda_{EWSB}$, the dimensionless effective Higgs mass squared, $\mu(\Lambda^2)$, is steep function rapidly approaching the second derivative of classical effective potential at its local minimum. This defines the physical Higgs mass (second derivative) and the EW VEV (local minimum). The $\Lambda_{EWSB}$ is less than an order of magnitude larger than VEV.

The "hierarchy problem" caused by quadratic running was traditionally utilized as motivation for new physics model building. The Supersymmetric models [6-10] address quadratic running with a set of supersymmetric partners (boson and fermions) whose leading quantum corrections to scalar mass squared exactly cancel. In models with dynamical symmetry breaking, like TechniColor (TC) [11-13], TopColor (TopC) [14], or TopColor assisted Technicolor (TC2) [15-23], there is no elementary Higgs scalar particle in ultraviolet; however, at some energy scales the running coupling constant is strong enough to form condensate.

Here, the cancelation of scalar quadratic divergences in the SM zero-VEV theory is promoted to self-consistent condition for the smooth dynamical EWSB phase transition within the composite framework. The scalar is a massive composite field in the infrared and non-existent in the ultraviolet, where effective scalar mass and scalar quartic coupling are both zero. This is similar to TechniColor theories where the "hierarchy problem" is avoided via dynamical EWSB. In difference to other dynamical models, the CPM has composite top quark and it is characterized by set of dynamical relationships. In the infrared the CPM mimics most of the features of an ordinary SM.

## 2. Review of the CPM

The minimally modified SM that can satisfy the cancelation of quadratic divergences in the zero-VEV theory has largest Yukawa coupling approximately equal to 1/3 of the SM top Yukawa coupling. The CPM [1-3] considers the massive top quark to have composite structure with 3 fundamental O quarks, $O \leftrightarrow \bar{O} \leftrightarrow O$, and the massive Higgs scalar is a color-neutral structure composed of 2 fundamental O quarks, $\bar{O} \leftrightarrow O$. Section 2.1 ref [3] analyzes the Higgs mass stability against radiative corrections for vanishing Higgs mass in the full SM calculation, with fixed, known, low energy SM gauge coupling values. The obtained condition for cancelation of leading divergences implies that the largest Yukawa coupling at low energies should be roughly equal to one third of the top quark Yukawa coupling. This motivates a CPM where massive top quark has composite structure with 3 original massless O quarks. Because O is not a mass eigenstate in the masssive EW phase, it cannot be observed directly.

Note that the SM gauge anomaly cancellation is satisfied as the elementary top quark is simply exchanged with the original O quark with identical SM gauge group couplings.

The three degrees of freedom (DoF) are gauged away to reappear as the three longitudinal polarizations of the W and Z massive gauge bosons whereas the forth DoF is represented by the physical Higgs particle. Hence, the fundamental O quark is not the mass eigenstate in the infrared and cannot be directly observed. But the true composite fermion, i.e. top quark, DoFs cannot be locally gauged away. Clearly, the heavy O particle with a mass between 50 and 60 GeV would be observed from W decays during searches for the heavy top quark.

Moreover, the CPM assumes radiative generation of the Higgs mass in the infrared and cancellation of surplus leading "divergences" in both 2D and 4D; here, the full SM renormalization scheme is applied to the running effective scalar mass squared. Section 2.2 ref [3] analyzes the possibility that Higgs mass (in the massive EW phase) is radiatively generated. If non-Abelian strong interactions are partially responsible for mass generation, then the 2D physical description might be appropriate toward the understanding of mass generation dynamics; the static charges observed in QCD form a spatially one dimensional tube. The obtained 2D equations lead to the prediction of Higgs mass parameterized by a positive integer k.

The two particularly interesting modes (k = 1, 2) are obtained and interpreted within the CPM [1-3]. The simplest CPM scenario in 2D leads to $m_H c^2$ = 113.0 GeV and 143.4 GeV corresponding to scalar masses affiliated with $\bar{O}O$ and $\bar{t}t$ condensates. These coincide with excesses observed by the Large Electron Positron (LEP) particle accelerator group [24,25] and by the Fermi National Accelerator Laboratory CDF collaboration [26].

As first proposed by Nambu in the simplest models with dynamical mass generation and fermion condensate in 4D one expects the Higgs mass on the order of twice the heaviest fermion mass [27,28]. If this is applied to the CPM one could expect that the top constituent subquarks and composite top quarks may generate scalar masses equal to $2\frac{m_t}{3}$ and $2m_t$ respectively.

As shown in Section 3, the $\bar{O}O$ and $\bar{t}t$ affiliated states (in both 2D and 4D representations) may mix in agreement with the Bose-Einstein statistics for $kT \cong M_W c^2$ which then lead to the physical Higgs mass $m_H \cong 125$ GeV (in both 2D and 4D representations) which coincide with excesses recently observed by the Large Hadron Collider (LHC) ATLAS [29-31] and CMS [32-35] collaborations.

The LEP group observed suspicious events [24,25] in the vicinity of 115 $GeV/c^2$, at the center of mass energies a bit above $\sqrt{s} \cong 206\ GeV/c^2$, just before the accelerator was shut down in 2000. Last year, the Fermi lab group reported excess in dijet invariant mass [26] centered in the vicinity of 144 $GeV/c^2$, at collision energies $\sqrt{s} \cong 1.96\ TeV/c^2$ just before the shutdown in September 2011. The LHC ATLAS [29-31] and CMS [32-35] collaborations recently announced 3 sigma level excess centered at 125 $GeV/c^2$.

Finally, the $\Lambda_{EWSB}$, Z, and W bosons masses are established via balanced top anti-top interactions at Z mass energy, i.e. via dynamical relationship presented in [2-3] between top Yukawa coupling and strong coupling constant $g_{QCD}$. In the language of ultraviolet physics, the O anti-O quarks will form a "lattice" at some energy scale (specifically Z mass). In the infrared language, this lattice will consist of nodes with top and anti-top quarks. Small deformation of the "lattice' will be represented by three Goldstone Bosons composites that are "eaten" by Z and W.

It is shown that the top anti-top scattering matrix at tree level is zero at the Z mass pole. In other words it takes zero energy to orient the top condensate field at the Z mass pole energy. This finding is supported by very accurate (0.3 %) prediction of the QCD gauge coupling at the Z pole from only the known value of the top quark mass.

Regarding potential completion of theory in the ultraviolet imagine that there is no fundamental scalar field in the high-energy sector of the theory. Instead, as one of several possible scenarios, imagine that there are non-SM four-fermion interactions that may be caused by, here unspecified, broken gauge symmetries at high energies. A heavy gauge boson G may be integrated out and the resulting contact terms may be Fiertz reordered to give the standard low-energy weak-interaction-like form of the non-SM four-fermion interaction terms. Furthermore, the four-fermion interaction terms may be interpreted as Yukawa term with an effective static scalar field, $\Phi$, i.e. field without a Lagrangian kinetic term. As shown before [36], the static effective scalar field may generate a fully gauge covariant kinetic term in the low energy effective theory via fermion loops. This can be summarized as

$$\bar{O}GO \leftrightarrow \bar{O}GO \xrightarrow{heavy\ G\ integrated\ out} \bar{O}O\bar{O}O \xrightarrow{static\ field\ \Phi} \bar{O}\Phi O. \qquad (1)$$

In the unbroken phase, the effective Lagrangian is

$$\mathcal{L}^{eff} = \mathcal{L}^{massless\ SM}_{(O\ instead\ of\ t)} + \mathcal{L}^{UV}_{(?)},\ and\ \langle\phi\rangle = 0\ . \qquad (2)$$

with theory in ultraviolet expressed at low energies by an unknown Lagrangian $\mathcal{L}^{UV}_{(?)}$ assumed to be irrelevant for calculations here. The $\mathcal{L}^{massless\ SM}_{(O\ instead\ of\ t)}$ term is a massless SM Lagrangian where top quark is exchanged with O quark having a different Yukawa coupling but identical SM gauge couplings as top quark,

$$\mathcal{L}^{massless\ SM}_{(O\ instead\ of\ t)} = \mathcal{L}^{SM}_{gauge\ bosons\ kinetic} + \mathcal{L}^{SM}_{fermions\ kinetic\ (O\ instead\ of\ t)} +$$

$$+\mathcal{L}^{SM}_{Yukawa\ (O\ instead\ of\ t)} + \begin{cases} \mathcal{L}^{SM}_{scalar\ kinetic} \\ 0 \end{cases}. \qquad (3)$$

As discussed in Sec. 2.1. ref [3] the identical CPM condition for cancelation of quadratic divergencies is obtained with and without scalar kinetic term.

In the broken phase the effective Lagrangian is identical to the SM Lagrangian to the extent that there may be more than one scalar field or dynamical resonances

$$\mathcal{L}^{eff} = \mathcal{L}^{massive\ SM}_{(t)} + \mathcal{L}_{massive\ scalar(s)}\ ,\ <\phi> \neq 0. \qquad (4)$$

### 2.1. The CPM Dynamical Relations

Here, the basic CPM dynamics relations are reviewed. For more detailed analysis see ref [3].

EWSB is a transition between two ground states. Here, instead of the SM Higgs and top quark fields in the unbroken massless phase, the CPM [1-3] has an auxiliary Higgs field with zero mass and quartic coupling, i.e. $m_H = 0$ and $\lambda = 0$, and an elementary quark field O instead of top quark, having Yukawa coupling equal to 1/3 of the SM top quark Yukawa coupling. As anticipated, for a $\langle\phi\rangle = 0$ ground state with SM gauge group, the minimally modified massless SM is stable against quadratic divergences at low energies, i.e. $\frac{dm_H^2(\Lambda^2)}{d\Lambda^2} \cong 0$, and the broken massive phase is expected to closely resemble SM with the O quark field confined within the Higgs $\bar{O} \leftrightarrow O$ and top quark $O \leftrightarrow \bar{O} \leftrightarrow O$ composite fields. In other words, the $\langle\phi\rangle \neq 0$ SM ground state is explained in terms of the original massless fields.

In the case of the SM scalar SU(2) Higgs doublet the three scalar degrees of freedom can be removed by the convenient gauge transformation. These degrees of freedom reappear as three Goldstone bosons that are "eaten" by Z and W; hence providing longitudinal polarization to massive gauge bosons. Similarly, in the case of O anti-O condensation, the convenient gauge transformation can remove the three O fermion degrees of freedom or equivalently the three composite scalar degrees of freedom whereas the fourth composite scalar degree of freedom is identified as composite scalar field similar to elementary Higgs.

Hence, it is not possible to directly observe O particle. Clearly, the heavy O particle with mass of ~ 60 GeV would be observed from W decays during searches for the top quark.

As shown in [3] the condition that leading SM scalar renormalized mass squared "divergences" cancels out in unbroken SM electroweak phase (i.e. zero VEV ground state) is identical with and without propagating Higgs in both 2D (see also [37, 2-3]) and 4D. By ignoring all other Yukawa couplings except the largest, one obtains

$$g_f^2 = \frac{g_Y^2 + 3g_W^2}{12} \text{ and } \lambda = 0. \tag{1}$$

For the low energy SM values of $U(1)_Y \times SU(2)_W$ gauge couplings $g_Y = \frac{e}{\cos\theta_W}, g_W = \frac{e}{\sin\theta_W}$ where $\theta_W$ is the weak mixing angle, one obtains

$$g_f \cong \frac{g_t}{3}. \tag{2}$$

No SM fermion has this Yukawa coupling and SM Higgs quartic coupling is not zero. However, in CPM this is exactly what one would expect in the unbroken phase!

As first emphasized by Nambu [27, 28], in simplest models with dynamical mass generation [39-41] and top condensate [27, 28, 42-44] EWSB, one may expect the Higgs mass on the order of 2 top quark masses. Hence, if top quark is exchanged with O quark one might expect the Higgs mass on the order of $\frac{2}{3}m_t$ [1-3].

The condition that radiatively generated scalar mass is in agreement with renormalization group equations in the 2D description is addressed next. As is demonstrated by the lattice arguments, e.g. see [38], the non-Abelian gauge fields carry charge that causes their propagation to mimic the 1-space D flux providing confinement between static charges. Hence, the 2D considerations here are thought of as consequence of non-Abelian gauge fields' dynamics in regular 4D.

As discussed in detail in [3], regarding the radiatively generated Higgs mass in 2D, the dominant fermion loop contribution to radiative corrections can be split into two terms: the first term (x-term) contributes to the cancellation of leading "divergences" in Equ (8) ref [3] and the second term (y-term) equals to the radiatively generated Higgs mass. Two terms add to one, i.e. $x + y = 1$ and the single color 2D fermion loop is exactly solved [2,3] as

$$2D \text{ fermion loop} = \frac{kg^2}{\pi}. \tag{3}$$

This result is obtained with similar techniques as the fermion loop in the Schwinger model [45]. However, in contrast to the Schwinger model this is a composite scalar and not a gauge boson. Here, an explicit dependence on the relevant phase space is parameterized with k. According to Schwinger model, the mass singularities in the propagator should exist for multiple integer values [38, 45], i.e. $k \in N$.

The obtained system has three unknowns, $x$, $y$ and $\lambda$, and three equations,

$$3\lambda + \frac{g_Y^2 + 3g_W^2}{4} = 3g_t^2 x, \quad x + y = 1 \text{ and } \lambda = \frac{kg_t^2}{\pi}y, \tag{4}$$

leading to an unique solution

$$3\lambda + \frac{g_Y^2 + 3g_W^2}{4} = 3g_t^2\left(1 - \frac{\lambda\pi}{kg_t^2}\right) \to \sqrt{\lambda} = \sqrt{\frac{g_t^2 - \frac{g_Y^2 + 3g_W^2}{12}}{1 + \frac{\pi}{k}}}, \tag{5}$$

$$\Rightarrow m_H = \sqrt{\frac{6m_t^2 - M_Z^2 - 2M_W^2}{3\left(1 + \frac{\pi}{k}\right)}}. \tag{6}$$

The above calculation, see also [2,3], is self consistent as the Higgs mass in the Higgs loop propagator (within piece proportional to $x$) is identical to radiatively generated Higgs mass (within piece proportional to $y$).

For the world average top quark mass, $m_t = 173.1 \pm 1.3 \, GeV$, one obtains
$$m_H = \begin{cases} 113.0 \pm 1.0 \, GeV \, for \, k = 1 \to y = \mathbf{0.669}, \, x = 0.331 \\ 143.4 \pm 1.3 \, GeV \, for \, k = 2 \to y = \mathbf{0.539}, \, x = 0.461 \end{cases}. \quad (7)$$

Interestingly, according to Equ (6), for the $k \to \infty$ the $m_H \to \cong 230 \, GeV$. This finding is compatible with the upper limit on the SM Higgs mass [5, 2] obtained from the requirement that the running effective Higgs mass takes zero value at any scale (which happens to be order $\sim TeV$ scale). The mapping between the physical SM Higgs mass and this scale, named Higgs Mass Zero Crossing (HMZC) scale, has been presented in [5, 2].

The Higgs mass squared in Equ (4) within 2D k=1 mode may be rewritten as
$$y \frac{g_t^2}{\pi} \cong \frac{2}{3} \frac{g_t^2}{\pi} = 6 \frac{(g_t/3)^2}{\pi} = n_S n_C \frac{(g_t/3)^2}{\pi} \quad (8)$$
where $n_S = 2$ and $n_C = 3$ are spin and color contributions to phase space respectively and where $\cong g_t/3$ is O Yukawa coupling. Hence, 2D k=1 Higgs appears to be generated via O anti-O condensate.

The Higgs mass squared in Equ (4) within 2D k=2 mode may be rewritten as
$$y \frac{2g_t^2}{\pi} \cong 0.539 \frac{2g_t^2}{\pi} = 0.539 \cdot 2 \cdot 9 \frac{(g_t/3)^2}{\pi} = 9.702 \frac{(g_t/3)^2}{\pi}. \quad (9)$$

This result may be represented as mixing between 4D top anti-top contribution to 2D dynamics and direct 2D O anti-O contribution. The top anti-top contribution to 2D k=2 Higgs is
$$n_S n_C \pi^2 \frac{(g_t/3)^2}{\pi} \quad (10)$$
where factor $\pi^2$ is the phase space of 2 transversal O anti-O condensates with half axial $(0, \pi)$ symmetry.

The weighted sum parameterized with an angle $\alpha$ should be equal to the 2D k = 2 mode Higgs mass. Hence,
$$\cos^2 \alpha \cdot n_S n_C + \sin^2 \alpha \cdot n_S n_C \pi^2 \cong 9.702 \quad \Rightarrow \quad \cos^2 \alpha = 0.930 \quad (11)$$

See ref [3] for review of 4D matching.

For completeness, the CPM condition defining weak scale is briefly reviewed next.

As discussed in ref [3] hypothesis that CPM is characterized by particular ground state such that top anti-top channel is neither attractive or repulsive at tree level at the Z pole mass correspond to
$$2 \frac{1}{2} \frac{2}{3} g_{QCD}^2 = g_t^2 \quad or \quad \alpha_S = \frac{3}{2} \frac{g_t^2}{4\pi} \quad at \, Z \, mass \quad (12)$$

The result in Equ (12) is in an excellent agreement with the standard estimate of the strong running coupling constant [47, 48]. Equ (12) predicts $\alpha_s = 0.1181 \pm 0.0018$ given the world average top quark mass $m_t = 173.1 \pm 1.3 \, GeV$ where uncertainty is therefore solely due to the top quark mass uncertainty. This can be compared with the world average value $\alpha_s \cong 0.1184 \pm 0.0007$ at $s = M_Z^2$ [47, 48].

## 3. The Bose-Einstein distribution applied to condensates mixing

Here, the possibility that Bose-Einstein statistics applied to the mixing of the lowest CPM energy states defines the Higgs mass provided that there is thermodynamics equilibrium as characteristics of the CPM ground state is analyzed next.

Imagine that Higgs field, $\Phi$, may be shared by O anti-O, top anti-top, and potentially something else

$$\Phi = c_{o\bar{o}} \Phi_{O\bar{O}} + c_{t\bar{t}} \Phi_{t\bar{t}} + \cdots \qquad (13)$$

where coefficients $c_{oo}, c_{tt}, \cdots$ are relative contributions normalized to one.

### 3.1. Mixing in 4D

By applying the Bose-Einstein distribution with assumed values

$$\hbar\omega_{o\bar{o}\,4D} \cong m_{o\bar{o}}c^2, \quad \hbar\omega_{t\bar{t}4D} \cong m_{t\bar{t}}c^2 \cong 3\,m_{o\bar{o}}c^2, \text{ and } kT \cong M_W c^2 \qquad (14)$$

with CPM's $m_{t\bar{t}} \cong 2m_t \cong 346.2$ GeV and $m_{O\bar{O}} \cong \frac{2}{3}m_t \cong 115.4$ GeV in 4D on the lines of top condensate models as emphasized by Nambu [27,44]. Hence,

$$\Rightarrow c_{o\bar{o}\,4D} = \frac{\frac{1}{e^{\hbar\omega_{oo\,4D}/kT}-1}}{\frac{1}{e^{\hbar\omega_{oo4D}/kT}-1} + \frac{1}{e^{\hbar\omega_{tt4D}/kT}-1} + \cdots} \leq 0.9585 \qquad (15)$$

The mass contribution to Higgs field from each condensate type can be expressed as

$$m_{H\,phys} = c_{o\bar{o}4D} \cdot 2\frac{m_t}{3} + c_{t\bar{t}4D} \cdot 6\frac{m_t}{3} + \cdots \qquad (16)$$

$$\Rightarrow m_{H\,phys} \geq 0.9585 \cdot 2\frac{m_t}{3} + (1 - 0.9585)6 \cdot \frac{m_t}{3} \cong 125.0\ GeV \qquad (17)$$

i.e. result in potentially excellent agreement with the recent excess obtained by the LHC [29-35].

### 3.2. Mixing in 2D

By applying the Bose-Einstein distribution with assumed values

$$\hbar\omega_{o\bar{o}\,2D} \cong 113.0\ GeV, \quad \hbar\omega_{t\bar{t}2D} \cong 143.4\ GeV, \text{ and } kT \cong M_W c^2 \qquad (18)$$

with CPM's k=1 and k=2 solutions in 2D, see Section 2.2.2. Hence,

$$\Rightarrow c_{o\bar{o}\,2D} = \frac{\frac{1}{e^{\hbar\omega_{oo\,2D}/kT}-1}}{\frac{1}{e^{\hbar\omega_{oo2D}/kT}-1} + \frac{1}{e^{\hbar\omega_{tt2D}/kT}-1} + \cdots} \leq 0.6168 \qquad (19)$$

The mass contribution to Higgs field from each condensate type can be expressed as

$$m_{H\,phys} = c_{o\bar{o}2D} \cdot 113.0\ GeV + c_{t\bar{t}2D} \cdot 143.4\ GeV + \cdots \qquad (20)$$

$$m_{H\,phys} \geq 0.6168 \cdot 113.0\ GeV + (1 - 0.6168)143.4\ GeV \cong 124.6\ GeV \qquad (21)$$

i.e. result in potentially excellent agreement with the recent excess obtained by the LHC [29-35].

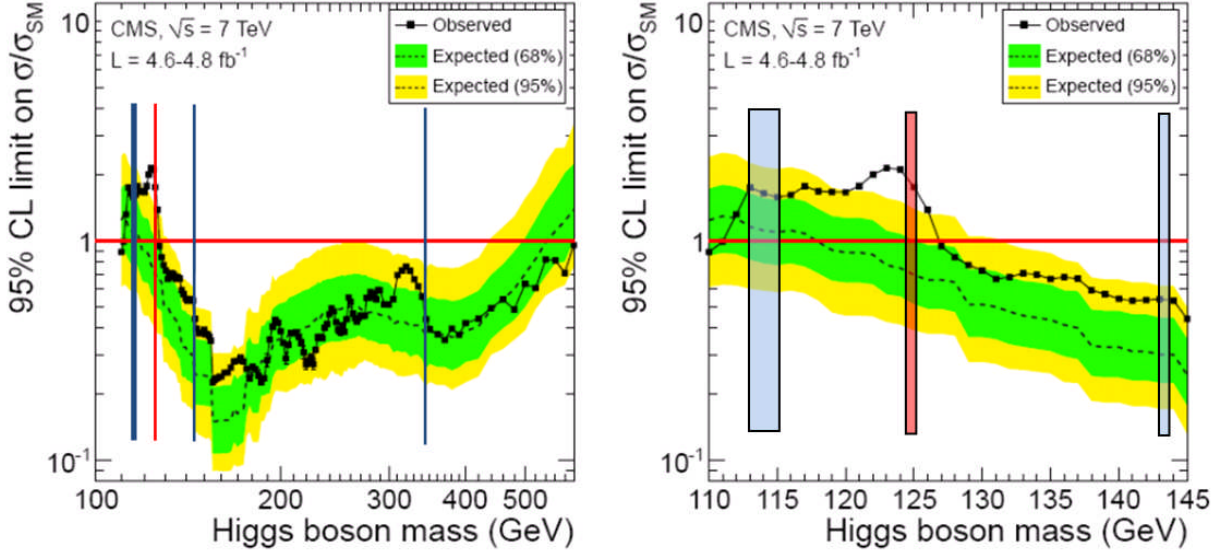

**Figure 1** The potential CPM scalar masses in 2D and 4D, vertical lines, are superimposed on the recent result obtained by the LHC CMS collaboration [34] showing the 95% CL upper limits on the signal strength parameter $\sigma/\sigma_{SM}$ for the SM Higgs boson hypothesis as a function of the Higgs boson mass.

## 4. Summary

The CPM [1-3] is reviewed and the concept of condensates mixing based on the Bose-Einstein statistics with $kT \cong M_W c^2$ is advanced.

The CPM provides natural explanation for mass generation via O anti-O and top anti-top interactions. The CMP proposed scalar masses provide potentially good fit to recent experimental observation by Fermi Lab [26] and CERN [29-35], see Figure 1.

In 2D the radiatively generated k=1 mode Higgs mass is expected to be $m_H = 113.0$ GeV whereas the k=2 mode mass is $m_H = 143.4$ GeV. These predictions are supported by self-consistent cancellation of extra leading "divergences".

As first proposed by Nambu, in the simplest models with dynamical mass generation and fermion condensate in 4D, one expects the Higgs mass on the order of twice the heaviest fermion mass. If this is applied to the CPM one could expect that the top's constituent subquarks and composite top quarks may generate scalar masses equal to $\frac{2}{3} m_t \cong 115.4$ GeV and $2m_t \cong 346.2$ GeV respectively.

When Bose-Einstein statistics for $kT \cong M_W c^2$ is applied to the two lowest energy states in 2D (113 GeV and 143 GeV) and 4D (115 GeV and 346 GeV), the CPM suggests physical Higgs mass equal to $m_H \cong 125$ GeV/$c^2$ in both 2D and 4D.

The hierarchy problem is resolved by non-existence of elementary Higgs in ultraviolet and by defining relationship Equ (12), [2,3]. This relationship between top Yukawa coupling and strong QCD coupling obtained by requiring that top anti-top channel are neither attractive or repulsive at tree level at $\sqrt{s} \cong M_Z c^2$, defines the Z mass and hence the EW VEV and the CPM Higgs mass.

Finally, Equ (12) defines the HMZC scale, [5, 2], at which the running effective Higgs mass squared is zero. As argued in [2] the HMZC $\cong$ EWSB scale if SM is the effective theory in vicinity of the EWSB scale, and if the Universe's dynamics were never dominantly tachyonic.

**Acknowledgement**

I am indebted to Gorazd Cvetic, Thomas Hambye, Dejan Stojkovic, and Chris T. Hill, for valuable comments and encouragements to publish earlier manuscripts [2,3]. I am thankful to everyone at the WPI Physics Department. I am indebted to Kristin Politi for editing part of this manuscript. I would also like to thank Kevin Vanslette for physics comments and editing.